\documentclass[10pt,conference]{IEEEtran}
\IEEEoverridecommandlockouts
\usepackage{booktabs} 
\usepackage{diagbox}
\usepackage{balance}

\usepackage{colortbl}
\usepackage{amsmath}

\usepackage{enumitem}
\usepackage{fancyvrb}
\usepackage{epstopdf}
\usepackage{color}
\usepackage{amsfonts}
\usepackage{listings}
\usepackage{graphicx}
\usepackage{algorithm}
\definecolor{darkblue}{rgb}{0.0, 0.0, 0.55}
\usepackage{verbatim}
\usepackage{multirow}
\usepackage{framed}
\usepackage{caption}
\DeclareCaptionType{copyrightbox}
\usepackage{subcaption}
\usepackage{pifont}
\usepackage{soul}
\usepackage{mdframed}

\usepackage[hyphens,spaces,obeyspaces]{url}  
\usepackage{hyperref}

\usepackage[font={small,bf}]{caption}
\usepackage{multicol}

\usepackage{fancyvrb}
\usepackage{soul}

\newcommand{\ignore}[1]{}
\definecolor{Light}{gray}{.85}

\usepackage{xcolor}
\usepackage{newverbs}

\newverbcommand{\cverb}{\color{red}}{}
\newverbcommand{\bverb}
  {\begin{lrbox}{\verbbox}}
  {\end{lrbox}\colorbox{gray!30}{\box\verbbox}}

\newcommand{\Name}{OLLM}

\usepackage{cite}
\usepackage{amsmath,amssymb,amsfonts}
\usepackage{algorithmic}
\usepackage{graphicx}
\usepackage{textcomp}
\usepackage{xcolor}
\def\BibTeX{{\rm B\kern-.05em{\sc i\kern-.025em b}\kern-.08em
    T\kern-.1667em\lower.7ex\hbox{E}\kern-.125emX}}
\begin{document}


\title{A Study of Using Multimodal LLMs for Non-Crash Functional Bug Detection in Android Apps}

\author{\IEEEauthorblockN{Bangyan Ju\IEEEauthorrefmark{1},
Jin Yang\IEEEauthorrefmark{1},
Tingting Yu\IEEEauthorrefmark{2},
Tamerlan Abdullayev\IEEEauthorrefmark{1},
Yuanyuan Wu\IEEEauthorrefmark{1},
Dingbang Wang\IEEEauthorrefmark{2},
Yu Zhao\IEEEauthorrefmark{1}}
\IEEEauthorblockA{\IEEEauthorrefmark{1}University of Cincinnati, USA; \\
}
\IEEEauthorblockA{\IEEEauthorrefmark{2}University of Connecticut, USA;\\
}
\IEEEauthorblockA{jubn@mail.uc.edu, yang3j7@mail.uc.edu, tingting.yu@uconn.edu, abdulltn@mail.uc.edu, wu3yy@mail.uc.edu\\
dingbang.wang@uconn.edu, zhao3y3@ucmail.uc.edu}}

\maketitle

\begin{abstract}
Numerous approaches employing various strategies have been developed to test the graphical user interfaces (GUIs) of mobile apps. However, traditional GUI testing techniques, such as random and model-based testing, primarily focus on generating test sequences that excel in achieving high code coverage but often fail to act as effective test oracles for non-crash functional (NCF) bug detection. To tackle these limitations, this study empirically investigates the capability of leveraging large language models (LLMs) to be test oracles to detect NCF bugs in Android apps. Our intuition is that the training corpora of LLMs, encompassing extensive mobile app usage and bug report descriptions, enable them with the domain knowledge relevant to NCF bug detection. We conducted a comprehensive empirical study to explore the effectiveness of LLMs as test oracles for detecting NCF bugs in Android apps on 71 well-documented NCF bugs. The results demonstrated that LLMs achieve a 49\% bug detection rate, outperforming existing tools for detecting NCF bugs in Android apps. Additionally, by leveraging LLMs to be test oracles, we successfully detected 24 previously unknown NCF bugs in 64 Android apps, with four of these bugs being confirmed or fixed. However, we also identified limitations of LLMs, primarily related to performance degradation, inherent randomness, and false positives. Our study highlights the potential of leveraging LLMs as test oracles for Android NCF bug detection and suggests directions for future research.
\end{abstract}


\section{Introduction}
\label{sec:intro}

Mobile applications (apps) have witnessed a surge in popularity, with the Google Play app store hosting approximately three million applications~\cite{googleplaylink}. A pivotal study revealed that a significant majority (88\%) of app users are likely to abandon an application if they consistently encounter functional problems~\cite{applause}. Indeed, apps that fail to function properly can have significant real-life consequences for users. This motivates developers to quickly identify and resolve issues, or risk losing users. Many automated GUI testing approaches for mobile apps have been proposed, such as random testing~\cite{monkey, machiry2013dynodroid}, model-based testing~\cite{amalfitano2012using, amalfitano2015mobiguitar, azim2013targeted}, and learning based testing~\cite{koroglu2018qbe, pan2020reinforcement, li2019humanoid, degott2019learning}. 
The goal of these automated test generators is to generate  event sequences to achieve high code
coverage and/or detect crashes. However, these tools mainly concentrate on designing test sequences instead of test oracles~\cite{lin2014accuracy} to verify the presence of bugs. As a result, these works typically identify only those bugs that cause a system to crash, which are evident from crash logs, rather than more subtle non-crash functional (NCF) bugs. The absence of mobile-specific testing oracles presents a significant challenge in detecting NCF bugs~\cite{linares2017continuous}. 

Some specialized tools have been developed to detect specific types of NCF bugs in limited scopes. For example, DiffDroid~\cite{fazzini2017automated} is a technique that automatically detects cross-platform inconsistencies in mobile apps. iFixDataloss~\cite{guo2022ifixdataloss} can detect data loss issues in Android apps. SetDroid~\cite{sun2021setdroid} can detect setting-related issues. All current methods are derived from limited observations and based on pre-defined heuristic rules (e.g., differential analysis~\cite{petsios2017nezha}) to detect a specific type of NCF bug with a dedicated test sequence that supports differential analysis. They can not generalize to a wide category of NCF bugs, accommodate general test sequences, and assure effectiveness. A recent study~\cite{xiong2023empirical} revealed that of 399 crawled NCF bugs from Github~\cite{githublink}, only 84 fall within the detection scope of seven of state-of-art existing tools~\cite{fazzini2017automated, guo2022ifixdataloss, sun2021setdroid, su2021fully, wang2022detecting, escobar2020empirical,su2021owleyes}, which identified merely two of them in total. Consequently, it is necessary to generate test oracles to detect NCF bugs with high accuracy in diverse categories. 

Our intuition in this study is that leveraging Multimodal Large Language Models (LLMs) as test oracles could extend the scope of detectable NCF bugs beyond the capabilities of existing tools and increase the successful detection rate. Multimodal LLMs~\cite{chowdhery2023palm, ozdemir2023quick}, such as GPT-4o~\cite{GPT4o} have significantly enhanced capabilities in natural language understanding, image processing, and question answering. By leveraging extensive, unlabeled text corpora and images for self-supervised learning, LLMs develop a deep reservoir of domain knowledge. For instance, GPT from OpenAI boasts billions of parameters and is trained on diverse datasets, including extensive mobile app usage and bug report descriptions. This comprehensive training equips LLMs with the domain knowledge essential for detecting NCF bugs effectively.



In this paper, we conduct an empirical study on using multimodal Large Language Models (LLMs) as test oracles for detecting NCF bugs. Different from a recent study~\cite{xiong2023empirical} that focuses on the causes of NCF functional bugs and the performance of existing tools, our research is dedicated to assessing the performance of LLMs as test oracles. We also provide insights and suggestions on the advantages and limitations of LLMs for future research. We have formulated four research questions to guide our study:

\noindent
{\bf RQ1:}
How effective and efficient are LLMs to be NCF bug test oracles compared to state-of-the-art bug detection tools?

\noindent
{\bf RQ2:}
What roles do two-prompt strategy, in-context learning, and rules in prompt play in the effectiveness of LLMs?

\noindent
{\bf RQ3:}
How do different models of LLMs, such as GPT and Gemini, perform in detecting NCF bugs?

\noindent
{\bf RQ4:}
What is the usability of using LLMs for detecting real-world NCF bugs with random test sequence generation?

To comprehensively study the capability of leveraging LLMs as test oracles for detecting NCF bugs in Android, we propose \Name{}, an attempt to leverage multimodal LLMs to be test oracles in NCF bug detection. \Name{} incorporates fundamental LLM mechanisms, including prompts and in-context learning~\cite{liu2023pre}. The input of \Name{} is a given test sequence represented by a sequence of GUI events and an APK of the app under test. As a proof-of-concept, we assume the test sequences are already generated by test generation approaches~\cite{amalfitano2012using, amalfitano2015mobiguitar, azim2013targeted, koroglu2018qbe, pan2020reinforcement, li2019humanoid, degott2019learning}, e.g, a random test sequence generation method used in section \ref{sec:real}. \Name{} includes three phases. In the first phase, \Name{} focuses on the collection of execution data during GUI testing. This involves monitoring and documenting essential information in the runtime testing, including the execution of events in the test sequence and the corresponding Execution Results (ERs). The ERs are represented as layout information that incorporates textual data and accompanying screenshot images supported by multimodal LLMs. The second phase involves designing two specific prompts as inputs for the LLM. One prompt is aimed at detecting UI logic errors using textual layout information, while the other prompt targets the identification of UI display issues by leveraging screenshot data. In the third phase, \Name{} employs the LLM to execute two prompts sequentially and integrates in-context learning strategies to enhance detection accuracy. Then \Name{} combines the insights from the two prompts to determine the presence of NCF bugs.

Utilizing GPT-4o as the LLM, our study results demonstrate that \Name{} has a broad NCF bug detection scope, covering all 71 well-documented Android NCF bugs in our dataset. In comparison, six state-of-the-art methods OwlEye~\cite{su2021owleyes}, iFixDataloss~\cite{guo2022ifixdataloss}, SetDroid~\cite{sun2021setdroid}, Genie~\cite{su2021fully}, Odin~\cite{wang2022detecting}, and ITDroid~\cite{escobar2020empirical} cover only 17, 2, 1, 2, 11, and 3 bugs, respectively.  \Name{} effectively identified 35 (49\%) of the NCF bugs and provided accurate descriptions of the detected bugs, significantly outperforming these existing tools. During applying LLMs, fundamental mechanisms, such as the two prompt strategy, in-context learning, and rule-based prompts, play significant roles in enhancing NCF bug detection. For instance, in-context learning alone increased the detection rate from 40\% to 49\%. By using five different models of LLMs, we observed significant performance differences among them. Our study was further extended to include a new dataset of 64 Android apps, in which \Name{} successfully uncovered 24 previously unrecognized bugs, four of which have been confirmed or fixed. Despite these successes, we also identified limitations of LLMs, primarily related to performance degradation, inherent randomness, and false positives. Our study underscores the potential of LLMs as test oracles for identifying NCF bugs and highlights areas for future research. 

The contributions of this paper are as follows:

\begin{itemize}[noitemsep,nolistsep]

\item 
Our empirical study comprehensively explores the performance of leveraging LLMs to be test oracles. The evaluation results demonstrate that LLMs are effective in detecting Android NCF bugs while also revealing certain limitations, and suggesting directions for future research.

\item 
Our research pioneers a novel attempt \Name{} by leveraging LLMs as test oracles for NCF bug detection in Android apps.


\item
\Name{} along with all experiment data are publicly available ~\cite{OLLM}.

\end{itemize}

\section{Preliminaries and Motivation}
\label{sec:pre}

\subsection{Preliminaries}
\label{sec:pre2}

Test sequences for mobile applications typically consist of sequences of Graphical User Interface (GUI) events\footnote{In our setting, an event refers to an executable GUI widget associated with
an action type (e.g., click, scroll, edit, swipe, etc). An action is defined as the execution of an event.}, designed to simulate user interactions with the apps. After executing an event, the application exhibits certain behaviors as Execution Results (ERs), such as navigating to a different page or displaying a pop-up dialog. A proficient test sequence effectively engages the appropriate GUI widget on the app screen and detects bugs that include crash and NCF bugs. Recent study~\cite{xiong2023empirical} shows 65.4\% of bugs are categorized as NCF bugs. In this work, we define NCF bugs as software issues that deviate from the expected functionalities of an operational app, without leading to an observable crash of the app,  aligning with existing research~\cite{xiong2023empirical, wang2022detecting, su2021fully}. We have excluded non-functional bugs, such as those related to performance or energy, as well as compatibility issues that occur only on specific devices.

\begin{figure*}[t]
 \centering
 \includegraphics[scale=0.41]{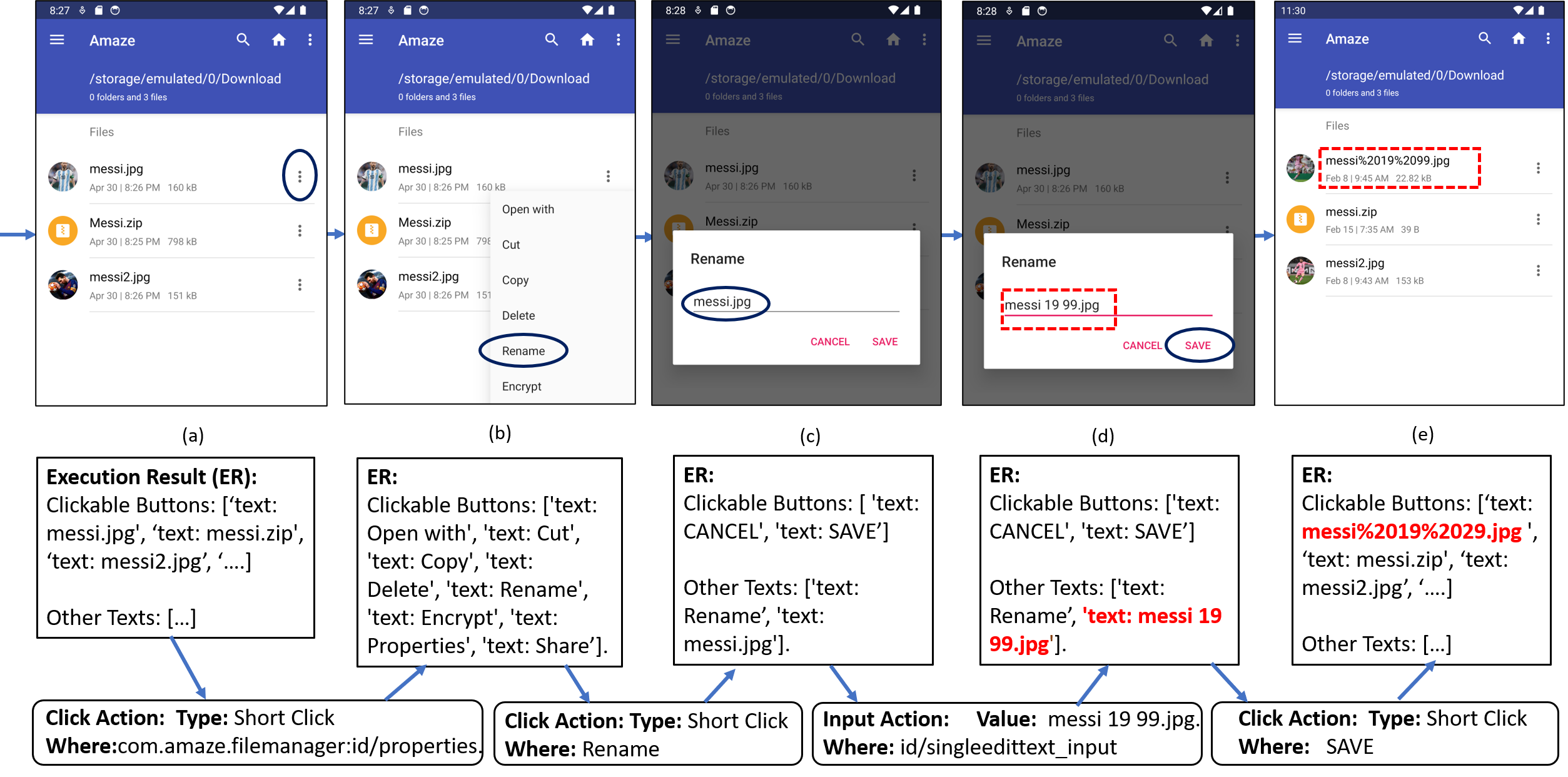}
  \caption{Motivation Example}
  \label{fig:mov}
  \vspace*{-20pt}
\end{figure*}

An illustration of a test sequence of Android bug Amaze-2113~\cite{Amaze-2113} is depicted in Fig.\ref{fig:mov}, which showcases a process of renaming a photo. In this scenario, a user attempts to rename a photo to ``Messi 19 99.jpg''. However, upon saving, the photo’s name is incorrectly displayed as ``messi\%2019\%2099.jpg'', deviating from the expected ``messi 19 99.jpg'', thereby revealing an NCF bug. In the depicted test sequence, ``Action" refers to the user interactions with the app, such as clicking and entering data. ``Execution Results" (ERs) display the textual resulting layouts after an action is performed.

\subsection{Observation}
\label{sec:cat.}
We studied hundreds of GitHub Android bug reports and drew on summarized categories based on empirical studies in existing works~\cite{xiong2023empirical,escobar2020empirical,guo2022ifixdataloss} to understand the characteristics of NCF bugs for guiding the design of \Name{} to detect these bugs. In general, NCF bugs can be grouped by UI logical bugs and UI display bugs that have been studied by existing work~\cite{xiong2023empirical}. \textbf{UI display bugs} can be categorized as follows: (1) Content-related issues (C.): the UI displays correctly but the content text does not make sense, such as truncated sentences; (2) Missing UI elements (M.): a UI component unexpectedly disappears; (3) UI distortion (UD.): displaying issues such as overlapping elements or blurry images; (4) Redundant UI elements (R.): UI components are duplicated. Similarly, \textbf{UI logical bugs} include: (1) Incorrect interaction (I.): the app's subsequent behavior deviates from expectations, such as clicking a button named ``category" redirects to the home screen; (2) Functionality does not take effect (F.): the user successfully interacts, but the function doesn't work; (3) UI element does not react (UR.): components do not react to user inputs, such as no reaction after clicking a button; (4) Data loss (D.): the data entered by the user is unexpectedly cleared; (5) Language related (L.): language issues in the international version of the app, such as unexpected language changes.

\subsection{Comparison with existing works}
\label{sec:existing}
Some existing works focus on detecting NCF bugs,  but all of them only focus on specific types of bugs rather than general NCF bugs. Several existing studies involve performing differential analyses of app statuses using dedicated test sequences. ITDroid~\cite{escobar2020empirical} can automate the detection of internationalization (i18n) issues, by doing differential testing using the original app and the internationalized one. SetDroid~\cite{sun2021understanding} is designed to identify issues related to user-configurable settings (e.g., network, location, permissions) that apps may fail to adapt properly. iFixDataloss~\cite{guo2022ifixdataloss} executes specific events, such as screen rotation, and then analyzes the differences in data on the layout before and after the event to identify data loss.  DiffDroid~\cite{fazzini2017automated} can automatically identify cross-platform inconsistencies in mobile applications. GENIE~\cite{su2021fully} leverages a differential analysis that utilizes a differential analysis to identify NCF bugs, which are characterized by changes in one part of an app that adversely affect other parts. ODIN~\cite{wang2022detecting} utilizes differential analysis to identify the abnormal behavior of one pair of GUI layouts compared to other pairs. The tools discussed are specifically designed to address a narrowly defined range of bug categories, relying on pre-defined heuristic rules based on limited observations. Additionally, they depend on dedicated test sequences that facilitate differential analysis, such as running two test sequences on different devices for DiffDroid. These limitations restrict their generalizability and effectiveness. Owleye~\cite{su2021owleyes} employs deep learning to model the visual information from GUI screenshots, focusing on identifying UI display issues such as text overlap and blurred screens. All these tools also lack the natural language understanding capacity to understand the abnormal behavior of apps. As noted in the introduction, a recent study~\cite{xiong2023empirical} reveals that only a small percentage of NCF bugs fall within their scope, with only about 0.5\% of these bugs being detectable. None of the existing tools are capable of successfully detecting the NCF bug illustrated in Fig. \ref{fig:mov}, which necessitates understanding the behavior of the text ``rename" both theoretically and practically. By incorporating extensive and in-depth domain knowledge, LLMs can detect a broader range of NCF bugs with heightened precision. 
\section{Proposed LLMs based Test Oracle}
\label{sec:approach}

As an attempt to leverage LLMs to be test oracles for NCF bug detection in this study, \Name{} should incorporate multiple essential components that have proven effective in several existing works~\cite{zhou2024large, wang2024feedback, liu2024make}. These components include data collection, prompt design, and in-context learning. The overarching structure of \Name{} encompasses three main phases: information extraction, prompt generation, and prompt execution. 

\subsection{Information Extraction from Test Sequence}
\label{sec:extraction}

\Name{} extracts both action information and corresponding ERs from the captured screen data of the runtime Android app undergoing testing sequences. The ERs encompass textual information and screenshot images from the layout supported by multimodal LLMs. 

The initial step of information extraction involves capturing runtime information during the execution of the test sequence. \Name{} dumps GUI layout information from an app following a series of user actions. The current version of \Name{} supports six types of UI actions: Click, Long Click, Drag, Swipe, and Input, as well as four system actions: Back, Home, Enter, and Rotate. During each testing iteration within a given test sequence, executing events such as ``Click Menu'', ``Click Rename", ``Input messi 19 99.jpg", and ``Click Save" as illustrated in Fig.\ref{fig:mov} trigger updates in the GUI pages, reflecting execution results (ERs) of the event execution. \Name{} employs UIAutomator~\cite{uiautomator} to perform these actions and then dump the textual screen layout and screenshot as the output of the corresponding action. 

\begin{table}[htbp]
\centering
\caption{\label{tab:text} Textual Extraction Results from (d) to (e) in Fig.\ref{fig:mov}}
\scalebox{0.9}{
\begin{tabular}{|l|l|}
\hline
Steps & Text Extraction                                                                                                                                          \\ \hline

(d)  & \begin{tabular}[c]{@{}l@{}}\textbf{Action}: Input 'id/singleedittext\_input' with 'messi 19 99.jpg'\\ \textbf{Clickable Buttons}: {[}'content-desc: CANCEL', \\ 'content-desc: SAVE’{]}\\ \textbf{Other Texts}: {[}'content-desc: Rename’, \\ 'content-desc: messi 19 99.jpg'{]}\end{tabular}                                                               \\ \hline
(e)  & \begin{tabular}[c]{@{}l@{}}\textbf{Action}: Click 'Save'\\ \textbf{Clickable Buttons}: {[}‘content-desc: messi\%2019\%2099.jpg', \\ ‘content-desc: messi.zip', ‘content-desc: messi2.jpg’, ‘…'{]}\\ \textbf{Other Texts}: {[}'content-desc: Amaze','content-desc:\\  /Storage/emulated/0/Download', ..{]}\end{tabular}                                                               \\ \hline
\end{tabular}
}
\vspace*{-5pt}
\end{table}

To effectively represent the ERs gleaned from the dumped information, \Name{} extracts essential data elements from the data dump of text layout including the text information on clickable buttons, long clickable buttons, checkable elements, and other texts. These other texts are defined as unexecutable text of the GUI, including labels, headings, and informational text. \Name{} organizes the extracted textual information from the test sequence into a structured format that includes actions and corresponding textual ERs. Table~\ref{tab:text} shows an example of the extracted textual information from steps (d) and (e) in the test sequence depicted in Fig.~\ref{fig:mov}.

In addition to text-based information extraction, \Name{} enhances its data extraction process by incorporating screenshot images corresponding to the action as an additional component of ERs. 




\begin{figure}[htbp]
\centering
\includegraphics[scale=0.5]{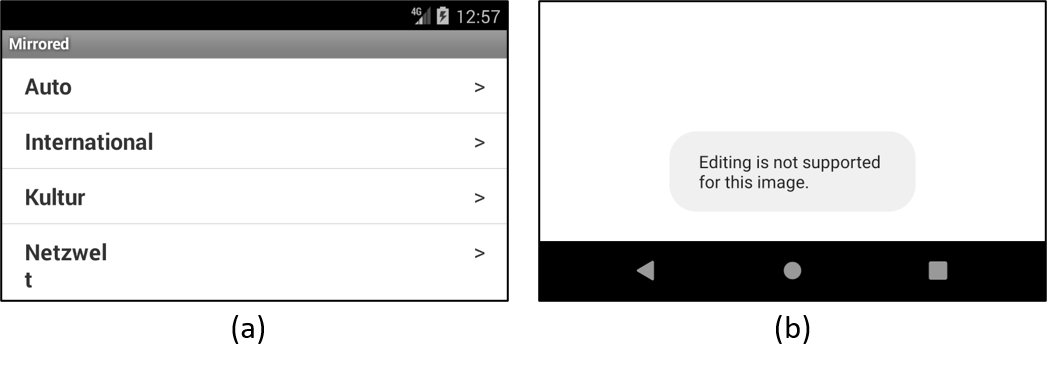}
\caption{Screenshots for Bug Detection}
\label{fig:image}
\vspace*{-20pt}
\end{figure}


\subsection{Two Prompt Generators}
\label{sec:multi-generators}
The second phase involves generating prompts that utilize information extracted from the execution of the test sequence. In the current version of \Name{}, we have adopted
the concept of Decomposed Prompting ~\cite{khot2022decomposed}, which breaks
down a complex prompt task into simpler sub-tasks. This
methodology optimizes each prompt for its respective subtasks. We have designed and implemented two prompt generators, one for UI logic bugs and one for UI display bugs, as detailed in Table~\ref{tab:instruction}. The first prompt (prompt 1) leverages only textual information from ERs and actions tailored for general NCF bugs. It can particularly detect logical bugs that can be identified through text, such as incorrect actions, calculations, decisions, or data processing. An example of an NCF bug targeted by prompt 1 is the text inconsistency bug shown in Fig.~\ref{fig:mov}. Information from actions and ERs from steps (a) through (d) provides essential context for understanding the entire scenario of the test sequence. All this information has been extracted from texts of ERs and actions as detailed in section~\ref{sec:extraction}. 

\begin{table*}[ht]
\centering
\caption{\label{tab:instruction} Prompt Structure}
\scalebox{0.95}{
\begin{tabular}{l l l}
\hline
\multicolumn{3}{|c|}{Prompt Structure}                                                                                                                                                 \\ \hline
\multicolumn{1}{|l|}{Category}                                                 & \multicolumn{1}{l|}{Prompt 1}                                                                                                                                         & \multicolumn{1}{l|}{Prompt 2}                                                                                                                \\ \hline
\multicolumn{1}{|l|}{Question}                                                 & \multicolumn{1}{l|}{\begin{tabular}[t]{@{}l@{}}You will read an event flow, containing page text \\ and actions to reach other pages. \\ Then I will ask you: Is there any logical error, \\ or a bug in the output after the given test sequence?\end{tabular}}                                                                                                                                                            & \multicolumn{1}{l|}{\begin{tabular}[t]{@{}l@{}}You will read an Android app event sequence, containing actions to\\  reach each page and the text on each page. \\ Then I will send you a screenshot of the last page. \\ Your task is to determine whether there is a UI error in the screenshot.\end{tabular}}                                                   \\ \hline
\multicolumn{1}{|l|}{Rules}                                                    & \multicolumn{1}{l|}{\begin{tabular}[t]{@{}l@{}} While evaluating logic errors, you should also \\ consider the rules below: (1) You should not analyze battery\\ ...\end{tabular}} & \multicolumn{1}{l|}{\begin{tabular}[t]{@{}l@{}}If you detect error messages, please also check the consistency \\ or correctness of these elements about the provided sequence. \\...\end{tabular}} \\ \hline
\multicolumn{1}{|l|}{\begin{tabular}[c]{@{}l@{}}Test  \\ sequence\end{tabular}} & \multicolumn{1}{l|}{Text Extraction Result in Table I}                                                                             & \multicolumn{1}{l|}{Text Extraction Result in Table I + Screenshot of the Last Page}                                                                                       \\ \hline
\multicolumn{1}{|l|}{\begin{tabular}[c]{@{}l@{}}Output \\ format\end{tabular}} & \multicolumn{1}{l|}{\begin{tabular}[c]{@{}l@{}}Provide your answer with yes or no. If your answer is yes, \\ please also provide the reason. \\...
\end{tabular}}                                                                                       & \multicolumn{1}{l|}{\begin{tabular}[c]{@{}l@{}}Provide your answer with yes or no. If your answer is yes, \\ please also provide the reason. \\
...\end{tabular}}                                                                                                \\ \hline
    &       &       \\
     &       &         \\
            &        &             \\
    &       &        
\end{tabular}
}
\vspace{-50 pt}
\end{table*}

The second prompt (prompt 2) incorporates both captured screenshot images and textual information from ERs and actions to target on UI display bug.  For example, a UI display bug illustrated in Fig.\ref{fig:image}(a) from the app Mirrored\cite{Mirrored} shows the text ``Netzwelt” being cut off at the bottom, with only the ``t" visible on the following line. Additionally, inconsistencies between dumped text information and UI display also need screenshot information to detect, as seen in the Android bug AnkiDroid-7232~\cite{AnkiDroid-7232} depicted in Fig.~\ref{fig:image}(b). The message ``Editing is not supported for this image'' can only be captured through screenshot images rather than text dumps. 

Our prompt incorporates LLMs with 4 detailed instruction sections ``Question'', ``Rules'', ``Test sequence'', and ``Output format'' in prompts to set clear expectations for the LLM's responses when detecting potential NCF bugs within an app. The Question section defines the focus for each prompt and claims the expectations in the LLM's response as if there is an NCF bug. The Rule section includes some constraints and hints that the LLM needs to follow. These rules are designed to minimize the false positives as the fake alarm. These constraints and hints are generated based on our extensive analysis of hundreds of bug reports and common errors made by LLMs. Detailed documentation of the 10 rules along with 4 instruction sections is available in our GitHub repository ~\cite{OLLM}.

Additionally, the extracted information in the test sequence, outlined in Section~\ref{sec:extraction}, forms a crucial part of the prompt as the third section of the Test Sequence. Specifically, to detect bugs that require visual cues in prompt 2, \Name{} integrates text extraction results with a screenshot of the final page. The decision to use only the last page's screenshot is due to the significant time and resource demands of image processing by these multimodal LLMs. Since \Name{} already incorporates test sequence information in text, including all screenshots is unnecessary. Using the unique image significantly reduces processing time. Lastly, the Output Format section is established to clarify the expected response format.

\subsection{Prompt Execution}
\label{sec:execution}


In the third phase, prompts are executed in the LLM to detect NCF bugs in responses. Since LLMs are typically not pre-trained for specific tasks like NCF bug detection, this may result in a low detection rate. To address this, recent research~\cite{liu2023pre} highlights the LLMs' ability to acquire new skills by learning from a few examples provided in the prompt, a process known as in-context learning. \Name{} employs in-context learning to enhance the effectiveness of NCF bug identification by providing question-answer pairs specifically curated from typical functional bug scenarios. In this study, these examples are randomly selected K-pairs from the corresponding group (logical or display bug) relevant to the prompt. We empirically set K to 3, consistent with the settings of recent work~\cite{deng2024mobile}.

\Name{} then aggregates responses from the two prompts to make a final decision regarding the presence of NCF bugs, a strategy we refer to as the integrated result of two prompts. To enhance the true positive of NCF bug detection, we have implemented a heuristic rule of the integrated result of two prompts: if either prompt suggests the presence of a bug, an NCF bug exists. Otherwise, there is no bug. 







\section{Study Design}
In this empirical study, we aim to study the comprehensive capability of LLMs to be NCF bug test oracles. Along with four research questions mentioned in section~\ref{sec:intro}, we also discuss the observed limitations of using LLMs, such as performance degradation, inherent randomness, and high false positive rates, and suggest directions for future research.

\subsection{Datasets}

To assess our methodology in RQ1, RQ2, and RQ3, it is crucial to establish datasets for evaluation. An existing study~\cite{xiong2023empirical} has constructed a dataset of NCF bugs sourced from GitHub with well-documented bug reports, employing a standard selection procedure. This dataset includes 399 NCF bugs and is aimed at evaluating current methodologies. Out of these, we retained 71 bugs identified through manual examination by our two graduate students as genuine NCF issues satisfying the definition in section ~\ref{sec:pre2}. This process involved approximately 400 hours of examining bug reports and manually writing test sequences to reach the bug. Bugs excluded from our evaluation include those with unavailable APKs, unclear reports, irreproducibility, or outdated apps. Additionally, requests for functional design improvements, like the date format request in WordPress-15026\cite{WordPress-15026}, are not considered bugs. As detailed in section ~\ref{sec:pre2}, performance and compatibility bugs are also omitted from our analysis.

To further validate the usability of \Name{} to detect unaware NCF bugs in RQ4, we applied it to a widely used app dataset that serves as the benchmark for various Android testing tools~\cite{mao2016sapienz, su2017guided}. This dataset comprises 68 apps spanning 18 domains~\cite{EvaluationDataSet}. We excluded four apps that consistently crashed immediately upon launch in our Intel Atom emulator. \Name{}'s objective differs from the existing tools that use this dataset primarily to identify crash bugs and enhance code coverage. 







\subsection{Evaluation Setup}
We conducted our experiment on a physical x86 machine equipped with an Intel i5 CPU and running Ubuntu 20.04. We primarily focused our investigation on GPT-4o~\cite{GPT4}, the most recent cutting-edge large language model from OpenAI to facilitate bug detection responses.

\textit{1) RQ1: Effectiveness and Efficiency of \Name{}}: To address Research Question 1, we assess \Name{} and six other Android functional detection tools (OwlEye~\cite{su2021owleyes}, iFixDataloss~\cite{guo2022ifixdataloss}, SetDroid~\cite{sun2021setdroid}, Genie~\cite{su2021fully}, Odin~\cite{wang2022detecting}, and ITDroid~\cite{escobar2020empirical}) in detecting NCF bugs and testing scope. We exclude DiffDroid~\cite{fazzini2017automated} from this comparison as it focuses on compatibility bugs, which are outside \Name{}'s scope. Since \Name{} does not generate test sequences, our analysis focuses solely on the effectiveness of test oracles using pre-defined test sequences. These test sequences used in this study are based on bug reports analyzed manually generated by two graduate students, ensuring they reach the bugs. We directly compare OwlEye and iFixDataloss with \Name{} using the same test sequences. However, the test oracles in SetDroid, Genie, Odin, and ITDroid require two specifically designed sequences to expose differences in application status and cannot operate on our manually generated sequences. To compare with these tools, we extract the testing scope of these tools by reviewing relevant papers and checking if their test oracles could detect the NCF bugs in our dataset, assuming these tools generated the correct inputs to trigger these bugs. Then we use the number of NCF bugs within the test scope of SetDroid, Genie, Odin, and ITDroid to represent the optimistic upper bound of their bug detection capability. An existing study~\cite{xiong2023empirical} has already measured the true positive rate of these tools on the same dataset, finding only 0.5\% of NCF bugs were detected, suggesting no need to reevaluate their overall performance.

False positives (FP) represent false alarms that waste developers' time during verification. To fairly assess false positives, we generate bug-free test sequences that closely resemble those used to evaluate true positives capable of reaching the bugs by removing the final step that triggered bugs. \Name{} was then tested on these modified sequences to determine its accuracy in confirming the absence of bugs. Furthermore, to enhance our evaluation beyond merely confirming bug presence, we introduced a metric called the True Positive Correct (TPC) rate that captures both the successful identification of bugs and the correctness of their descriptions as specified in the bug documentation. In the following text, ``detected bug by \Name{}" indicates that the bug meets the TPC metric criteria. In this study, we calculate the TPC rate by dividing the number of true positive correct results by the total number of actual positives. Conversely, the FP rate is calculated as the ratio of the number of negative events wrongly categorized as positive to the total number of actual negative cases.

We evaluate \Name{} across configurations to assess its overall performance including the default setting of querying once with the integrated result of two prompts (detailed in section~\ref{sec:execution}). We evaluate \Name{} in two modes: querying once (q1) and querying three times (q3) using the same prompt. The q3 mode is employed to thoroughly examine the LLM's performance by minimizing its randomness, as it offers \Name{} three opportunities to provide the correct response. If at least one TPC appears in the results of the three queries, we regard it as a successful TPC detection under q3 mode. Conversely, if at least one FP occurs in a bug-free sequence, we categorize it as an FP alarm. For each mode, we separately repeat the settings using only prompt 1, only prompt 2, and the integrated result of two prompts, totaling six settings.



\textit{2) RQ2: Roles of two-prompt strategy, in-context Learning, and rules in prompt} within \Name{}, we investigate how different components influence its effectiveness and efficiency for RQ2. To do this, we repeated the experiment with the two-prompt strategy, in-context learning, prompt rules components disabled individually, then reevaluated \Name{}'s performance under q1 mode. The performance of prompt 1 and prompt 2 are analyzed exclusively. \Name{}-PA merges and consolidates all unduplicated contents from prompts 1 and 2 by disabling the prompt decomposition strategy outlined in section~\ref{sec:multi-generators}. OLLM-PA is then used to analyze the case presented by a single prompt in a complex task. \Name{}-NoCon is utilized to assess \Name{} by disabling in-context learning, as described in Section \ref{sec:execution}. \Name{}-NoRule is used to evaluate the performance impact of disabling the ``Rules'' section, which serves as constraints and guidance in the prompts, as outlined in Table~\ref{tab:instruction}. 


\textit{3) RQ3: Performance of Different Models of LLMs:}
We assess \Name{} across two leading multimodal large language models that facilitate image processing, including Gemini~\cite{Gemini}, and ChatGLM-4~\cite{ChatGLM-4}, along with other GPT versions (GPT-3.5, 4~\cite{GPT4}), to assess their capabilities under q1 mode. 

\textit{4) RQ4: Usability on real-world previously-unaware apps}: To rigorously assess \Name{}'s effectiveness in identifying new bugs in real-world applications, we developed an automated test sequence generator utilizing a semi-random strategy. Unlike Monkey~\cite{monkey}, which employs a fully random strategy to select events, our method randomly selects one of the least frequently visited events to execute on the current page of the app. This strategy, inspired by Stoat~\cite{su2017guided}, aims to increase code coverage. Each selected event in the test sequence is executed as an action and dumped layout information is recorded, which \Name{} then uses to detect NCF bugs. The testing duration is set to one hour per app, aligning with the benchmarks set by other Android app testing tools~\cite{mao2016sapienz, li2019humanoid, pan2020reinforcement, vuong2018reinforcement}.

\section{Results and Analysis}
\subsection{RQ1: Effectiveness and Efficiency of detecting NCF bugs}
Table \ref{tab:result} presents the results of NCF bug detection by \Name{} across 71 bugs in 6 different settings. In the 'Total' and 'Percentage' rows, values before the parentheses represent results under the single-query mode (q1), and those within parentheses correspond to the three-query mode (q3). By using the default setting of the integrated result of two prompts under q1, \Name{} achieves a TPC rate of 49\% reflecting the accurate identifying of actual NCF bugs, and a false positive (FP) rate of 59\% indicating the rate of false alarm. The setting with the lowest FP rate is under mode q1 with unique prompt 1: TPC rate=42\% and FP rate= 29\%. This setting is ideal for test oracles prioritizing bug-free test results while maintaining a reasonable NCF bug detection rate. Conversely, the setting with the highest TPC rate involves three queries with the integrated result from two prompts: TPC rate =63\% and FP rate=93\%. This setting is suitable for test oracles that aim to reveal as many NCF bugs as possible but may suffer from a high FP rate. These results underscore \Name{}'s capability to successfully detect NCF bugs with accurate descriptions of the detected NCF bugs.

\begin{table}[h!]
\centering
\caption{\label{tab:result} Test Results of \Name{}}
\scalebox{0.775}{
\begin{tabular}{|l|l|l|ll|ll|ll|}
\hline
\multirow{2}{*}{Bugs} & \multirow{2}{*}{TC} & \multirow{2}{*}{Cat.} & \multicolumn{2}{c|}{Prompt 1}        & \multicolumn{2}{c|}{Prompt 2}        & \multicolumn{2}{c|}{Integrated result} \\ \cline{4-9} 
                      &                     &                       & \multicolumn{1}{l|}{TPC}    & FP     & \multicolumn{1}{l|}{TPC}    & FP     & \multicolumn{1}{l|}{TPC}    & FP     \\ \hline
AnkiDroid-7465        & 2                   & I.                    & \multicolumn{1}{l|}{ }   &  \textcolor{blue}{\checkmark}1   & \multicolumn{1}{l|}{ }   & \hspace{0.5em} 1  & \multicolumn{1}{l|}{ }   &  \textcolor{blue}{\checkmark}2   \\ \hline
AnkiDroid-7232        & 2                   & F.                    & \multicolumn{1}{l|}{ \textcolor{blue}{\checkmark}2}   &     & \multicolumn{1}{l|}{ \textcolor{blue}{\checkmark}3}   & \hspace{0.5em} 1   & \multicolumn{1}{l|}{ \textcolor{blue}{\checkmark}5}   & \hspace{0.5em} 1   \\ \hline
AnkiDroid-7836        & 4                   & UD.                   & \multicolumn{1}{l|}{ }   &  \textcolor{blue}{\checkmark}1   & \multicolumn{1}{l|}{ }   &  \textcolor{blue}{\checkmark}3   & \multicolumn{1}{l|}{ }   &  \textcolor{blue}{\checkmark}4   \\ \hline
AnkiDroid-7801        & 4                   & F.                    & \multicolumn{1}{l|}{ \textcolor{blue}{\checkmark}1}   &     & \multicolumn{1}{l|}{ }   &     & \multicolumn{1}{l|}{ \textcolor{blue}{\checkmark}1}   &     \\ \hline
AnkiDroid-7793        & 2                   & F.                    & \multicolumn{1}{l|}{ }   & \hspace{0.5em} 1   & \multicolumn{1}{l|}{ }   & \hspace{0.5em} 1   & \multicolumn{1}{l|}{ }   & \hspace{0.5em} 2   \\ \hline
AnkiDroid-7768        & 6                   & F.                    & \multicolumn{1}{l|}{\hspace{0.5em} 1}   &     & \multicolumn{1}{l|}{\hspace{0.5em} 2}   &  \textcolor{blue}{\checkmark}1   & \multicolumn{1}{l|}{\hspace{0.5em} 3}   &  \textcolor{blue}{\checkmark}1   \\ \hline
AnkiDroid-7730        & 7                   & I.                    & \multicolumn{1}{l|}{ }   &     & \multicolumn{1}{l|}{ }   &  \textcolor{blue}{\checkmark}2   & \multicolumn{1}{l|}{ }   &  \textcolor{blue}{\checkmark}2   \\ \hline
AnkiDroid-7674        & 2                   & F.                    & \multicolumn{1}{l|}{ \textcolor{blue}{\checkmark}2}   & \hspace{0.5em} 1   & \multicolumn{1}{l|}{ }   &  \textcolor{blue}{\checkmark}2   & \multicolumn{1}{l|}{ \textcolor{blue}{\checkmark}2}   &  \textcolor{blue}{\checkmark}3   \\ \hline
AnkiDroid-6288        & 2                   & F.                    & \multicolumn{1}{l|}{ \textcolor{blue}{\checkmark}3}   & \hspace{0.5em} 1   & \multicolumn{1}{l|}{ \textcolor{blue}{\checkmark}3}   &     & \multicolumn{1}{l|}{ \textcolor{blue}{\checkmark}6}   & \hspace{0.5em} 1   \\ \hline
AnkiDroid-7377        & 4                   & UD.                   & \multicolumn{1}{l|}{ }   &  \textcolor{blue}{\checkmark}2   & \multicolumn{1}{l|}{ }   &  \textcolor{blue}{\checkmark}1   & \multicolumn{1}{l|}{ }   &  \textcolor{blue}{\checkmark}3   \\ \hline
AnkiDroid-7070        & 4                   & F.                    & \multicolumn{1}{l|}{\hspace{0.5em} 2}   &     & \multicolumn{1}{l|}{ }   &  \textcolor{blue}{\checkmark}2   & \multicolumn{1}{l|}{\hspace{0.5em} 2}   &  \textcolor{blue}{\checkmark}2   \\ \hline
AnkiDroid-7027        & 4                   & R.                    & \multicolumn{1}{l|}{ }   & \hspace{0.5em} 1   & \multicolumn{1}{l|}{ \textcolor{blue}{\checkmark}3}   & \hspace{0.5em} 2   & \multicolumn{1}{l|}{ \textcolor{blue}{\checkmark}3}   & \hspace{0.5em} 3   \\ \hline
AnkiDroid-6887        & 8                   & I.                    & \multicolumn{1}{l|}{ }   & \hspace{0.5em} 2   & \multicolumn{1}{l|}{ }   &  \textcolor{blue}{\checkmark}2   & \multicolumn{1}{l|}{ }   &  \textcolor{blue}{\checkmark}4   \\ \hline
AnkiDroid-6894        & 3                   & I.                    & \multicolumn{1}{l|}{\hspace{0.5em} 1}   &     & \multicolumn{1}{l|}{ }   & \hspace{0.5em} 1   & \multicolumn{1}{l|}{\hspace{0.5em} 1}   & \hspace{0.5em} 1   \\ \hline
AnkiDroid-5688        & 2                   & I.                    & \multicolumn{1}{l|}{\hspace{0.5em} 1}   & \hspace{0.5em} 1   & \multicolumn{1}{l|}{ }   &  \textcolor{blue}{\checkmark}1   & \multicolumn{1}{l|}{\hspace{0.5em} 1}   &  \textcolor{blue}{\checkmark}2   \\ \hline
AnkiDroid-5091        & 4                   & I.                    & \multicolumn{1}{l|}{ \textcolor{blue}{\checkmark}3}   & \hspace{0.5em} 1   & \multicolumn{1}{l|}{ }   &  \textcolor{blue}{\checkmark}1   & \multicolumn{1}{l|}{ \textcolor{blue}{\checkmark}3}   &  \textcolor{blue}{\checkmark}2   \\ \hline
AnkiDroid-5167        & 4                   & F.                    & \multicolumn{1}{l|}{ }   & \hspace{0.5em} 1   & \multicolumn{1}{l|}{ }   & \hspace{0.5em} 1   & \multicolumn{1}{l|}{ }   & \hspace{0.5em} 2   \\ \hline
AnkiDroid-4935        & 0                   & I.                    & \multicolumn{1}{l|}{ \textcolor{blue}{\checkmark}1}   &     & \multicolumn{1}{l|}{ }   & \hspace{0.5em} 1   & \multicolumn{1}{l|}{ \textcolor{blue}{\checkmark}1}   & \hspace{0.5em} 1   \\ \hline
AnkiDroid-8975        & 2                   & F.                    & \multicolumn{1}{l|}{ }   &     & \multicolumn{1}{l|}{ }   & \hspace{0.5em} 1   & \multicolumn{1}{l|}{ }   & \hspace{0.5em} 1   \\ \hline
AnkiDroid-8379        & 3                   & I.                    & \multicolumn{1}{l|}{ \textcolor{blue}{\checkmark}2}   &  \textcolor{blue}{\checkmark}1   & \multicolumn{1}{l|}{ }   &  \textcolor{blue}{\checkmark}1   & \multicolumn{1}{l|}{ \textcolor{blue}{\checkmark}2}   &  \textcolor{blue}{\checkmark}2   \\ \hline
AnkiDroid-7023        & 1                   & F.                    & \multicolumn{1}{l|}{ }   &  \textcolor{blue}{\checkmark}1   & \multicolumn{1}{l|}{ }   & \hspace{0.5em} 1   & \multicolumn{1}{l|}{ }   &  \textcolor{blue}{\checkmark}2   \\ \hline
AntennaPod-4776       & 2                   & I.                    & \multicolumn{1}{l|}{ \textcolor{blue}{\checkmark}3}   & \hspace{0.5em} 1   & \multicolumn{1}{l|}{ \textcolor{blue}{\checkmark}2}   &  \textcolor{blue}{\checkmark}3   & \multicolumn{1}{l|}{ \textcolor{blue}{\checkmark}5}   &  \textcolor{blue}{\checkmark}4   \\ \hline
AntennaPod-3786       & 2                   & I.                    & \multicolumn{1}{l|}{ \textcolor{blue}{\checkmark}1}   & \hspace{0.5em} 1   & \multicolumn{1}{l|}{ }   &     & \multicolumn{1}{l|}{ \textcolor{blue}{\checkmark}1}   & \hspace{0.5em} 1   \\ \hline
AntennaPod-2992       & 0                   & M.                    & \multicolumn{1}{l|}{ }   &  \textcolor{blue}{\checkmark}1   & \multicolumn{1}{l|}{\hspace{0.5em} 1}   &  \textcolor{blue}{\checkmark}3   & \multicolumn{1}{l|}{\hspace{0.5em} 1}   &  \textcolor{blue}{\checkmark}4   \\ \hline
AntennaPod-4548       & 2                   & I.                    & \multicolumn{1}{l|}{ \textcolor{blue}{\checkmark}3}   & \hspace{0.5em} 1   & \multicolumn{1}{l|}{ }   & \hspace{0.5em} 2   & \multicolumn{1}{l|}{ \textcolor{blue}{\checkmark}3}   & \hspace{0.5em} 3   \\ \hline
Firefox-3617          & 2                   & F.                    & \multicolumn{1}{l|}{\hspace{0.5em} 1}   & \hspace{0.5em} 1   & \multicolumn{1}{l|}{ }   & \hspace{0.5em} 1   & \multicolumn{1}{l|}{\hspace{0.5em} 1}   & \hspace{0.5em} 2   \\ \hline
Firefox-3152          & 0                   & D.                    & \multicolumn{1}{l|}{ \textcolor{blue}{\checkmark}1}   &     & \multicolumn{1}{l|}{ }   &     & \multicolumn{1}{l|}{ \textcolor{blue}{\checkmark}1}   &     \\ \hline
Firefox-3291          & 3                   & I.                    & \multicolumn{1}{l|}{ }   & \hspace{0.5em} 1   & \multicolumn{1}{l|}{ }   & \hspace{0.5em} 1   & \multicolumn{1}{l|}{ }   & \hspace{0.5em} 2   \\ \hline
Firefox-4068          & 3                   & C.                    & \multicolumn{1}{l|}{ }   &     & \multicolumn{1}{l|}{ }   &     & \multicolumn{1}{l|}{ }   &     \\ \hline
Firefox-3146          & 4                   & L.                    & \multicolumn{1}{l|}{ \textcolor{blue}{\checkmark}2}   &  \textcolor{blue}{\checkmark}1   & \multicolumn{1}{l|}{ \textcolor{blue}{\checkmark}2}   & \hspace{0.5em} 2   & \multicolumn{1}{l|}{ \textcolor{blue}{\checkmark}4}   &  \textcolor{blue}{\checkmark}3   \\ \hline
Firefox-3254          & 2                   & F.                    & \multicolumn{1}{l|}{ \textcolor{blue}{\checkmark}2}   &  \textcolor{blue}{\checkmark}2   & \multicolumn{1}{l|}{ \textcolor{blue}{\checkmark}3}   &     & \multicolumn{1}{l|}{ \textcolor{blue}{\checkmark}5}   &  \textcolor{blue}{\checkmark}2   \\ \hline
Firefox-3121          & 1                   & F.                    & \multicolumn{1}{l|}{ }   &  \textcolor{blue}{\checkmark}2   & \multicolumn{1}{l|}{ }   &  \textcolor{blue}{\checkmark}1   & \multicolumn{1}{l|}{ }   &  \textcolor{blue}{\checkmark}3   \\ \hline
Firefox-3304          & 1                   & C.                    & \multicolumn{1}{l|}{ }   & \hspace{0.5em} 1   & \multicolumn{1}{l|}{ }   &     & \multicolumn{1}{l|}{ }   & \hspace{0.5em} 1   \\ \hline
Firefox-3297          & 2                   & D.                    & \multicolumn{1}{l|}{ }   &     & \multicolumn{1}{l|}{ }   & \hspace{0.5em} 1   & \multicolumn{1}{l|}{ }   & \hspace{0.5em} 1   \\ \hline
Simplenote-1294       & 3                   & F.                    & \multicolumn{1}{l|}{\hspace{0.5em} 2}   & \hspace{0.5em} 1   & \multicolumn{1}{l|}{ }   &  \textcolor{blue}{\checkmark}2   & \multicolumn{1}{l|}{\hspace{0.5em} 2}   &  \textcolor{blue}{\checkmark}3   \\ \hline
Simplenote-1190       & 1                   & F.                    & \multicolumn{1}{l|}{\hspace{0.5em} 2}   &     & \multicolumn{1}{l|}{ \textcolor{blue}{\checkmark}2}   &  \textcolor{blue}{\checkmark}1   & \multicolumn{1}{l|}{ \textcolor{blue}{\checkmark}4}   &  \textcolor{blue}{\checkmark}1   \\ \hline
Simplenote-1111       & 3                   & I.                    & \multicolumn{1}{l|}{ \textcolor{blue}{\checkmark}3}   &  \textcolor{blue}{\checkmark}3   & \multicolumn{1}{l|}{ \textcolor{blue}{\checkmark}3}   &     & \multicolumn{1}{l|}{ \textcolor{blue}{\checkmark}6}   &  \textcolor{blue}{\checkmark}3   \\ \hline
Simplenote-1046       & 2                   & UD.                   & \multicolumn{1}{l|}{ }   &     & \multicolumn{1}{l|}{ }   &     & \multicolumn{1}{l|}{ }   &     \\ \hline
Simplenote-984        & 1                   & UD.                   & \multicolumn{1}{l|}{ \textcolor{blue}{\checkmark}3}   &     & \multicolumn{1}{l|}{ }   &  \textcolor{blue}{\checkmark}2   & \multicolumn{1}{l|}{ \textcolor{blue}{\checkmark}3}   &  \textcolor{blue}{\checkmark}2   \\ \hline
Simplenote-952        & 1                   & UD.                   & \multicolumn{1}{l|}{ }   &  \textcolor{blue}{\checkmark}1   & \multicolumn{1}{l|}{\hspace{0.5em} 2}   &  \textcolor{blue}{\checkmark}2   & \multicolumn{1}{l|}{\hspace{0.5em} 2 }   &  \textcolor{blue}{\checkmark}3   \\ \hline
Simplenote-623        & 0                   & I.                    & \multicolumn{1}{l|}{ }   & \hspace{0.5em} 1   & \multicolumn{1}{l|}{ }   & \hspace{0.5em} 2   & \multicolumn{1}{l|}{}   & \hspace{0.5em} 3   \\ \hline
AnkiDroid-7758        & 5                   & L.                    & \multicolumn{1}{l|}{ }   & \hspace{0.5em} 1   & \multicolumn{1}{l|}{\hspace{0.5em} 1}   &  \textcolor{blue}{\checkmark}3   & \multicolumn{1}{l|}{\hspace{0.5em} 1}   &  \textcolor{blue}{\checkmark}4   \\ \hline
AnkiDroid-6857        & 0                   & I.                    & \multicolumn{1}{l|}{ }   &     & \multicolumn{1}{l|}{ }   & \hspace{0.5em} 2   & \multicolumn{1}{l|}{ }   & \hspace{0.5em} 2   \\ \hline
AnkiDroid-7366        & 1                   & UD.                   & \multicolumn{1}{l|}{ }   &     & \multicolumn{1}{l|}{ \textcolor{blue}{\checkmark}3}   &     & \multicolumn{1}{l|}{ \textcolor{blue}{\checkmark}3}   &     \\ \hline
AnkiDroid-6587        & 9                   & I.                    & \multicolumn{1}{l|}{ \textcolor{blue}{\checkmark}3}   &  \textcolor{blue}{\checkmark}2   & \multicolumn{1}{l|}{ \textcolor{blue}{\checkmark}3}   &  \textcolor{blue}{\checkmark}1   & \multicolumn{1}{l|}{ \textcolor{blue}{\checkmark}6}   &  \textcolor{blue}{\checkmark}3   \\ \hline
AnkiDroid-6119        & 2                   & I.                    & \multicolumn{1}{l|}{ }   &     & \multicolumn{1}{l|}{ }   & \hspace{0.5em} 1   & \multicolumn{1}{l|}{ }   & \hspace{0.5em} 1   \\ \hline
AnkiDroid-5334        & 5                   & I.                    & \multicolumn{1}{l|}{ \textcolor{blue}{\checkmark}3}   &  \textcolor{blue}{\checkmark}2   & \multicolumn{1}{l|}{ \textcolor{blue}{\checkmark}3}   &  \textcolor{blue}{\checkmark}2   & \multicolumn{1}{l|}{ \textcolor{blue}{\checkmark}6}   &  \textcolor{blue}{\checkmark}4   \\ \hline
AnkiDroid-5156        & 2                   & M.                    & \multicolumn{1}{l|}{ }   &     & \multicolumn{1}{l|}{ }   &  \textcolor{blue}{\checkmark}2   & \multicolumn{1}{l|}{ }   &  \textcolor{blue}{\checkmark}2   \\ \hline
AnkiDroid-5105        & 1                   & UD.                   & \multicolumn{1}{l|}{ }   &     & \multicolumn{1}{l|}{ }   & \hspace{0.5em} 2   & \multicolumn{1}{l|}{ }   & \hspace{0.5em} 2   \\ \hline
AnkiDroid-4999        & 3                   & I.                    & \multicolumn{1}{l|}{ \textcolor{blue}{\checkmark}1}   &     & \multicolumn{1}{l|}{ }   &  \textcolor{blue}{\checkmark}1   & \multicolumn{1}{l|}{ \textcolor{blue}{\checkmark}1}   &  \textcolor{blue}{\checkmark}1   \\ \hline
AnkiDroid-8072        & 5                   & F.                    & \multicolumn{1}{l|}{ \textcolor{blue}{\checkmark}3}   &  \textcolor{blue}{\checkmark}1   & \multicolumn{1}{l|}{ \textcolor{blue}{\checkmark}3}   &  \textcolor{blue}{\checkmark}1   & \multicolumn{1}{l|}{ \textcolor{blue}{\checkmark}6}   &  \textcolor{blue}{\checkmark}2   \\ \hline
AnkiDroid-8466        & 2                   & F.                    & \multicolumn{1}{l|}{ }   &  \textcolor{blue}{\checkmark}1   & \multicolumn{1}{l|}{ }   &  \textcolor{blue}{\checkmark}2   & \multicolumn{1}{l|}{ }   &  \textcolor{blue}{\checkmark}3   \\ \hline
AnkiDroid-8547        & 3                   & I.                    & \multicolumn{1}{l|}{ \textcolor{blue}{\checkmark}1}   &     & \multicolumn{1}{l|}{ }   &  \textcolor{blue}{\checkmark}2   & \multicolumn{1}{l|}{ \textcolor{blue}{\checkmark}1}   &  \textcolor{blue}{\checkmark}2   \\ \hline
AnkiDroid-7896        & 3                   & C.                    & \multicolumn{1}{l|}{ }   &  \textcolor{blue}{\checkmark}1   & \multicolumn{1}{l|}{ \textcolor{blue}{\checkmark}3}   & \hspace{0.5em} 2   & \multicolumn{1}{l|}{ \textcolor{blue}{\checkmark}3}   &  \textcolor{blue}{\checkmark}3   \\ \hline
Amaze-2113            & 2                   & I.                    & \multicolumn{1}{l|}{ \textcolor{blue}{\checkmark}3}   &  \textcolor{blue}{\checkmark}1   & \multicolumn{1}{l|}{ \textcolor{blue}{\checkmark}3}   &  \textcolor{blue}{\checkmark}1   & \multicolumn{1}{l|}{ \textcolor{blue}{\checkmark}6}   &  \textcolor{blue}{\checkmark}2   \\ \hline
Amaze-1919            & 1                   & I.                    & \multicolumn{1}{l|}{ \textcolor{blue}{\checkmark}3}   &     & \multicolumn{1}{l|}{ \textcolor{blue}{\checkmark}3}   &  \textcolor{blue}{\checkmark}2   & \multicolumn{1}{l|}{ \textcolor{blue}{\checkmark}6}   &  \textcolor{blue}{\checkmark}2   \\ \hline
Amaze-1916            & 3                   & I.                    & \multicolumn{1}{l|}{\hspace{0.5em} 2}   & \hspace{0.5em} 1   & \multicolumn{1}{l|}{ \textcolor{blue}{\checkmark}3}   & \hspace{0.5em} 1   & \multicolumn{1}{l|}{ \textcolor{blue}{\checkmark}5}   & \hspace{0.5em} 2   \\ \hline
Amaze-1872            & 3                   & UR.                   & \multicolumn{1}{l|}{ \textcolor{blue}{\checkmark}3}   &  \textcolor{blue}{\checkmark}1   & \multicolumn{1}{l|}{ }   & \hspace{0.5em} 1   & \multicolumn{1}{l|}{ \textcolor{blue}{\checkmark}3}   &  \textcolor{blue}{\checkmark}2   \\ \hline
Amaze-1834            & 3                   & I.                    & \multicolumn{1}{l|}{ \textcolor{blue}{\checkmark}3}   &     & \multicolumn{1}{l|}{ }   & \hspace{0.5em} 1   & \multicolumn{1}{l|}{ \textcolor{blue}{\checkmark}3}   & \hspace{0.5em} 1   \\ \hline
Amaze-1797            & 2                   & I.                    & \multicolumn{1}{l|}{ \textcolor{blue}{\checkmark}3}   & \hspace{0.5em} 2   & \multicolumn{1}{l|}{\hspace{0.5em} 2}   & \hspace{0.5em} 2   & \multicolumn{1}{l|}{ \textcolor{blue}{\checkmark}5}   & \hspace{0.5em} 4   \\ \hline
Amaze-1712            & 7                   & F.                    & \multicolumn{1}{l|}{ \textcolor{blue}{\checkmark}3}   &     & \multicolumn{1}{l|}{ }   & \hspace{0.5em} 1   & \multicolumn{1}{l|}{ \textcolor{blue}{\checkmark}3}   & \hspace{0.5em} 1   \\ \hline
Amaze-1628            & 2                   & F.                    & \multicolumn{1}{l|}{ \textcolor{blue}{\checkmark}2}   &     & \multicolumn{1}{l|}{ }   & \hspace{0.5em} 2   & \multicolumn{1}{l|}{ \textcolor{blue}{\checkmark}2}   & \hspace{0.5em} 2   \\ \hline
NewPipe-5363          & 4                   & I.                    & \multicolumn{1}{l|}{ }   & \hspace{0.5em} 1   & \multicolumn{1}{l|}{\hspace{0.5em} 2}   & \hspace{0.5em} 1   & \multicolumn{1}{l|}{\hspace{0.5em} 2}   & \hspace{0.5em} 2   \\ \hline
NewPipe-6409          & 1                   & C.                    & \multicolumn{1}{l|}{ }   &     & \multicolumn{1}{l|}{ }   &  \textcolor{blue}{\checkmark}2   & \multicolumn{1}{l|}{ }   &  \textcolor{blue}{\checkmark}2   \\ \hline
NewPipe-4113          & 3                   & M.                    & \multicolumn{1}{l|}{ }   &     & \multicolumn{1}{l|}{ }   &  \textcolor{blue}{\checkmark}3   & \multicolumn{1}{l|}{ }   &  \textcolor{blue}{\checkmark}3   \\ \hline
NewPipe-6397          & 2                   & I.                    & \multicolumn{1}{l|}{ \textcolor{blue}{\checkmark}2}   &  \textcolor{blue}{\checkmark}1   & \multicolumn{1}{l|}{ \textcolor{blue}{\checkmark}3}   & \hspace{0.5em} 2   & \multicolumn{1}{l|}{ \textcolor{blue}{\checkmark}5}   &  \textcolor{blue}{\checkmark}3   \\ \hline
WordPress-14234       & 4                   & I.                    & \multicolumn{1}{l|}{ }   & \hspace{0.5em} 1   & \multicolumn{1}{l|}{ }   &  \textcolor{blue}{\checkmark}3   & \multicolumn{1}{l|}{ }   &  \textcolor{blue}{\checkmark}4   \\ \hline
WordPress-13671       & 2                   & I.                    & \multicolumn{1}{l|}{ }   &  \textcolor{blue}{\checkmark}1   & \multicolumn{1}{l|}{ }   &  \textcolor{blue}{\checkmark}3   & \multicolumn{1}{l|}{ }   &  \textcolor{blue}{\checkmark}4   \\ \hline
WordPress-13121       & 3                   & UD.                   & \multicolumn{1}{l|}{ }   &     & \multicolumn{1}{l|}{ }   &  \textcolor{blue}{\checkmark}3   & \multicolumn{1}{l|}{ }   &  \textcolor{blue}{\checkmark}3   \\ \hline
WordPress-9966        & 3                   & I.                    & \multicolumn{1}{l|}{ \textcolor{blue}{\checkmark}3}   &  \textcolor{blue}{\checkmark}1   & \multicolumn{1}{l|}{ }   &  \textcolor{blue}{\checkmark}2   & \multicolumn{1}{l|}{ \textcolor{blue}{\checkmark}3}   &  \textcolor{blue}{\checkmark}3   \\ \hline
WordPress-8755        & 4                   & M.                    & \multicolumn{1}{l|}{ \textcolor{blue}{\checkmark}1}   & \hspace{0.5em} 1   & \multicolumn{1}{l|}{ \textcolor{blue}{\checkmark}3}   &  \textcolor{blue}{\checkmark}3   & \multicolumn{1}{l|}{ \textcolor{blue}{\checkmark}4}   &  \textcolor{blue}{\checkmark}4   \\ \hline Total: 71
                      & 2.7                 &                       & \multicolumn{1}{l|}{30(38)} & 21(43) & \multicolumn{1}{l|}{19(24)} & 34(61) & \multicolumn{1}{l|}{35(45)} & 42(66) \\ \hline  Percentage: 
                      &                     &                       & \multicolumn{1}{l|}{\begin{tabular}[c]{@{}l@{}}42\%\\ (53\%)\end{tabular}} & \begin{tabular}[c]{@{}l@{}}29\%\\ (61\%)\end{tabular} & \multicolumn{1}{l|}{\begin{tabular}[c]{@{}l@{}}26\%\\ (34\%)\end{tabular}} & \begin{tabular}[c]{@{}l@{}}48\%\\ (86\%)\end{tabular} & \multicolumn{1}{l|}{\begin{tabular}[c]{@{}l@{}}49\%\\ (63\%)\end{tabular}} & \begin{tabular}[c]{@{}l@{}}59\%\\ (93\%)\end{tabular} \\ \hline
\end{tabular}
}
\begin{flushleft}
Integrated result= The integrated result of two prompts
TC= number of texts required for comprehension to detect the bug.
Cat.= category,
TPC= true positive correct,
FP= false positive,
\checkmark= TPC/FP in first time query (q1),
Number n in TPC/FP column = number of TPC/FP in 3 queries (q3),
M(N)= number of first time query(number of at least one occurrence in 3 queries (q3))
\end{flushleft}
\vspace*{-30pt}
\end{table}

\Name{} and the six existing works are designed to detect NCF bugs. However, due to the limitations of heuristic-based techniques of existing works, \Name{} achieves much broader bug detection scopes as discussed in section ~\ref{sec:existing}. With pre-trained in-depth domain knowledge of LLM and generated prompts, \Name{} targets a wider array of NCF bug categories. Currently, we are not aware of any of the 71 NCF bugs in our dataset that fall outside \Name{}'s detection scope. In contrast, most bugs fall outside the designated scopes of existing tools. OWLEYE, IFixDataloss, ITDroid, SetDroid, Genie, and Odin cover only 17, 2, 3, 1, 2, and 11 bugs respectively. This limitation is a primary reason for their inability to detect these bugs. For example, in our experiments with the same dataset, \Name{} correctly identified and described 38 NCF bugs, whereas OwlEye and iFixDataloss detected only 4 and none, respectively. ITDroid, SetDroid, Genie, and Odin were able to detect up to 3, 1, 2, and 11 bugs, respectively. \Name{} utilizes an LLM pre-trained model on an extensive textual corpus, enabling it to generalize and detect a wide range of NCF bugs with higher accuracy.

The exceptional TPC rate of \Name{} is attributed to its proficiency in semantic understanding of text via LLMs. Analysis of 71 bugs in the dataset revealed that 66 bugs heavily depend on comprehending the actions and layout texts within the test sequence to detect the bug. The average number of instances requiring text semantic understanding for bug detection in these test sequences is 2.7 as shown in the TC column of Table~\ref{tab:result}. As illustrated by the example in Fig.~\ref{fig:mov}, the essential text comprehension of ``Files", ``Rename" and ``Save" provide \Name{} with sufficient information for logical reasoning and comparing two text strings: ``messi 19 99.jpg" and ``messi\%2019\%2029.jpg," thereby enabling \Name{} to detect NCF bugs. In contrast, previous approaches ~\cite{fazzini2017automated, guo2022ifixdataloss, sun2021setdroid, su2021fully, wang2022detecting, escobar2020empirical} relying solely on pre-defined heuristic rules and differential comparison. These methods depended entirely on specific differential analysis-based test sequences that could reveal differences in the status of Android apps for comparison. Without leveraging NLP techniques these methods fail to identify general functional issues. With the capacity of semantic understanding, \Name{} achieves a remarkable bug TPC rate by effectively understanding the semantic text information of runtime data in the test sequence without depending on differential analysis-based test sequences.

\subsection{RQ2: Effectiveness and Efficiency of Two-prompt strategy, In-context Learning, and Rules in Prompt}
\label{sec:config}

\begin{table}[]
\caption{\label{tab:MultipleLLMs} Multiple LLMS}
\scalebox{0.8}{
\begin{tabular}{|l|ll|ll|ll|}
\hline
          & \multicolumn{2}{l|}{Prompt1}                & \multicolumn{2}{l|}{Prompt2}               & \multicolumn{2}{l|}{Integrated result}                 \\ \hline
          & \multicolumn{1}{l|}{TPC}        & FP        & \multicolumn{1}{l|}{TPC}       & FP        & \multicolumn{1}{l|}{TPC}       & FP        \\ \hline
GPT-4o    & \multicolumn{1}{l|}{30 (42\%)}  & 21 (29\%) & \multicolumn{1}{l|}{19 (26\%)} & 34 (48\%) & \multicolumn{1}{l|}{35 (49\%)} & 42 (59\%) \\ \hline
GPT-4     & \multicolumn{1}{l|}{27 (38\%)}  & 56 (79\%)          & \multicolumn{1}{l|}{23 (32\%)}          & 48 (68\%)          & \multicolumn{1}{l|}{37 (52\%)}          &63  (89\%)           \\ \hline
GPT-3.5   & \multicolumn{1}{l|}{0  \hspace{0.2em} (0\%)}  & 45 (63\%)          & \multicolumn{1}{l|}{-}          &-           & \multicolumn{1}{l|}{0  \hspace{0.2em}  (0\%)}          & 45 (63\%)           \\ \hline
Gemini    & \multicolumn{1}{l|}{21 (30\%)}  & 12 (17\%)          & \multicolumn{1}{l|}{6  \hspace{0.2em} (8\%)}          & 44 (62\%)         & \multicolumn{1}{l|}{22 (31\%)}          & 48  (68\%)          \\ \hline
ChatGLM-4 & \multicolumn{1}{l|}{9   \hspace{0.2em} (13\%)} & 42 (59\%)         & \multicolumn{1}{l|}{10 (14\%)}          & 14 (20\%)         & \multicolumn{1}{l|}{16 (23\%)}          & 48  (68\%)           \\ \hline
\end{tabular}
}
\vspace*{-20pt}
\end{table}

Within \Name{}, we assess the performance of two prompts separately. The results indicate that prompt 1 achieved a 42\% TPC rate and a 29\% FP rate, whereas prompt 2 achieved a 26\% TPC rate and a 48\% FP rate. Prompt 1 successfully detects 62\% more NCF bugs and 40\% fewer mistakes in incorrectly identifying bug-free sequences as buggy compared to prompt 2, indicating prompt 2 is distracted by additional screenshot image information with text and overlooks some detailed text. Additionally, we find the average processing time of prompt 2 is 7.5 seconds which is much higher than prompt 1 which is 3 seconds. However, prompt 2 is able to detect 5 bugs that prompt 1 cannot, with three of these bugs falling into the UI display bug group as in section~\ref{sec:cat.}, including one instance of redundant UI elements, one instance of UI distortion, and one instance of content related issue. These findings confirm our expectation that Prompt 2, with its visual cues, is particularly effective at detecting specific NCF bugs, especially those related to UI display. OLLM-PA with a 35\% TPC rate performs worse than OLLM, which has a 49\% TPC rate under the same setting, though both share an FP rate of 59\%. This highlights that managing complex tasks with a single prompt is less effective than employing a two-prompt decomposed approach in detecting NCF bugs.


By removing the in-context learning, the results reveal that \Name{}-NoCon exhibited a lower TPC rate of 40\%, compared to the 49\% TPC rate observed with \Name{} equipped with in-context learning. These findings suggest that the inclusion of in-context learning in \Name{} plays a critical role in increasing the TPC rate. The results reveal that \Name{}-NoRule on prompt 1 exhibited a significantly high FP rate of 93\%, substantially greater than the 29\% observed with the original \Name{} on prompt 1 under mode q1. These findings suggest that the inclusion of rules in \Name{} plays a critical role in reducing the FP rates.

\subsection{RQ3: Performance of Multiple LLMs}
Overall, with reference to Table~\ref{tab:MultipleLLMs}, we find GPT-4 achieves the highest TPC rate at 52\% but it suffers the highest FP rate at 89\%. GPT-4o achieves a slightly lower TPC rate at 49\% but has the lowest FP rate at 42\%. However, GPT 3.5 and ChatGLM-4 exhibit significantly lower TPC rates of 0\% and 16\%. GPT-3.5's TPC rate is 0\% because all detected bugs have incorrect bug descriptions, violating the definition of TPC. Additionally, GPT-3.5 cannot query on prompt 2 as it does not support image input.  This study highlights the substantial performance differences among various LLMs when used as test oracles for NCF bug detection. Based on our findings, we recommend utilizing GPT-4o due to its strong TPC rate and lower FP rate. 

\subsection{RQ4: Real World NCF bug Detection with Test Sequence Generation}
\label{sec:real}


\Name{} effectively detected 24 NCF bugs in five bug categories across 64 Android applications included in the dataset. These bugs are unaware by the authors and not reported by literature~\cite{mao2016sapienz, su2017guided, zhao2022dinodroid} using the same dataset. One graduate student manually verifies the bug reported by \Name{} to ensure these bugs satisfy the NCF bug definition in section~\ref{sec:pre2}. 
Two issues have been confirmed and three of them have been resolved. In this context, ``confirmed" means that the bug has been reported and acknowledged by the developer, while ``fixed" refers to bugs that have been resolved by developers in subsequent versions. 
Table~\ref{tab:real} illustrates a selection of the NCF bugs detected by \Name{}, including the names of the affected apps, the categories of the bugs, and brief descriptions of each issue. \Name{} demonstrates the ability to detect NCF bugs across various categories by simply generating test sequences randomly. In addition to the successful detection rate, we also calculate the FP rate with randomly picked 25 randomly generated bug-free sequences in the dataset. The FP rates for prompt 1 and prompt 2 are 32\% and 52\%. Detailed reports of these findings are accessible in our experimental dataset~\cite{OLLM}.


\subsection{Limitations in LLMs and Future Research Direction}
\label{sec:real}
The most significant limitation observed in this study is the performance degradation of online business LLMs over time. Our study conducted in May 2024, using GPT-4 achieved a TPC rate of 43.7\% with old version of prompt 1 of \Name{}. We repeat the study with the same prompt and setting in July 2024, the result shows the TPC rate plummeted to 1\%.  Due to this serious performance degradation, we were compelled to modify our prompt design and incorporate an in-context learning strategy to achieve an acceptable detection rate, as demonstrated in this study. This severe instability in LLM performance could significantly impact the practical application of LLMs as test oracles. A potential solution is to design an adaptive learning system for prompt generation and in-context learning that can utilize machine learning to adjust its settings corresponding the current performance of the LLM.




We also observed significant randomness in the performance of LLMs. As indicated in Table~\ref{tab:result}, the TPCs of prompt 1 and prompt 2 under mode q3 are 26\% and 31\% higher than the TPCs detected in the first query. The increase in TPCs with repeated queries highlights the randomness in LLM responses. In 38 TPC cases under q3 in prompt 1, 22 cases missed at least one successful detection across the three queries. This level of randomness results in a significant time cost in manually verifying results, as the initial query may not yield an acceptable outcome due to this variability. A potential solution involves leveraging the BLEU metric~\cite{callison2006re} to identify common elements in the responses across repeated queries. The most frequent of these common elements in multiple queries can then represent the prime decision of LLMs and be used as the bug detection result. 


The FP rate of LLMs remains high, as evidenced in Table III. Even with the setting that yields the lowest FP rate, only prompt 1 under q1 mode, the rate is still 29\% in this study. Pursuing a higher TPC rate leads to an FP rate of  93\%, which is unacceptable. This high rate of FP substantially increases the time users spend on verifying results from LLMs. A potential solution could be to develop a machine learning model specifically designed to filter out some of these FP cases. Additionally, training a dedicated LLM on NCF bug related documentation, such as bug reports, could optimize the model to address these limitations more effectively.

Finally, OLLM lacks a dedicated mechanism for event selection or the design of test sequences. To improve the effectiveness of NCF bug testing in Android apps, a specialized test sequence generation method is needed.

\begin{table}[]
\caption{\label{tab:real} Selected Real World NCF Bug Detection}
\scalebox{0.8}{
\begin{tabular}{|l|l|l|l|l|}
\hline
App Name     & App Category        & \begin{tabular}[c]{@{}l@{}}Bug \\ Cat.\end{tabular} & Bug Description                                                                     & Status    \\ \hline

{Adsdroid}     & Reference & F.                                                  & Unable to fetch any results.                                                        & \href{https://github.com/dnet/adsdroid/issues/4}{\textbf{{Confirmed}}}+Fixed     \\ \hline
Mileage      & Utility             & C.                                                  & Displays unrealistic values.                                                        & \href{https://github.com/evancharlton/android-mileage/issues/47}{\textbf{{Confirmed}}}    \\ \hline

Wikipedia    & Reference & F.                                                  & Pages cannot be saved.                                                              & Fixed     \\ \hline
Anymemo      & Educational         & F.                                                  & Link directs to invalid site.                                                       & Fixed     \\ \hline
Anymemo      & Educational         & I.                                                  & Look up in dictionary fail.                                                                 & \href{https://github.com/helloworld1/AnyMemo/issues/535}{\textbf{{Reported}}} \\ \hline
Manpages    & Educational       & I.                                                  & All manual pages not found.                       & Manually verified     \\ \hline
Mirrored     & News                & F.                                                  & Articles cannot load.                                                               &  Manually verified         \\ \hline
Tippytipper  & Utility Tool        & F.                                                  & Tip counting incorrectly.                                                           &  Manually verified         \\ \hline
Filexplorer    & File Manager          & UR.                                                  & Buttons are non-responsive.                                                        &  Manually verified         \\ \hline
Weight-Chart        & Health         & UD.                                                  & Text overlays on grid lines.                                                          &  Manually verified         \\ \hline
yahtzee      & Game             & C.                                                  & Message contradicts text.                                                              & Manually verified     \\ \hline
\end{tabular}
}
\vspace*{-15pt}
\end{table}



\section{Related work}

Automated Android GUI testing has developed significantly, utilizing a variety of methods. Simple approaches like random testing~\cite{monkey} often lead to the generation of redundant and ineffective events. Tools such as DynoDroid~\cite{machiry2013dynodroid} aim to enhance efficiency by minimizing this redundancy, yet they still fall short in rigorously testing comprehensive app functionalities. Meanwhile, model-based strategies~\cite{amalfitano2012using, gu2019practical, amalfitano2015mobiguitar, azim2013targeted, baek2016automated, gu2017aimdroid, yan2018land, yan2017widget, yang2013grey} rely on constructing a GUI model of the app to inform test generation, often employing finite state machines to delineate app states and transitions. For instance, Stoat~\cite{su2017guided} uses a stochastic Finite State Machine model to simulate the app's behavior under test. Machine learning-based testing~\cite{koroglu2018qbe, borges2018guiding, li2019humanoid, lin2019test, degott2019learning} utilizes machine learning and deep learning techniques to produce test sequences that probe for Android crashes. Both QBE and DinoDroid~\cite{koroglu2018qbe, zhao2022dinodroid} apply reinforcement learning to derive testing methods for Android apps from a training set. However, these methodologies predominantly concentrate on generating test sequences to detect crash bugs and do not address the detection of NCF bugs. 

As outlined in Section~\ref{sec:existing}, some research focuses on identifying NCF bugs in Android applications. However, these studies predominantly target specific bug types within a narrow scope, rather than addressing general NCF bugs. These methods have demonstrated a modest success rate, detecting only about 2 in 399 NCF bugs when the dataset consists of bugs crawled from open-source apps on GitHub, according to a recent empirical study~\cite{xiong2023empirical}. Beyond the NCF bug detection tool discussed in Section~\ref{sec:existing}, other approaches also work on NCF bugs. For instance, AppFlow~\cite{hu2018appflow} and AppTestMigrator~\cite{behrang2019test} reuse test sequences in similar apps to validate functional correctness. Augusto~\cite{mariani2018augusto} and FARLEAD-Android~\cite{koroglu2021functional} utilize manually defined app functionalities in specification languages to detect NCF bugs. Tools like KREfinder~\cite{shan2016finding} and LiveDroid~\cite{farooq2020livedroid} employ static analysis to scan the source code for specific issues related to inconsistent states of Android apps. Compared to \Name{}, these tools require human input or source code access to identify NCF bugs. \Name{} only needs Android Apks and test sequences generated by any of the automated Android  GUI testing tools as mentioned in the last paragraph.

Research in software engineering, beyond NCF bug detection, is increasingly leveraging LLMs to enhance performance. For instance, adbGPT~\cite{feng2024prompting} utilizes an LLM to replicate Android bugs. Similarly, PG-TD~\cite{zhang2023planning} employs LLMs to generate code. GPTDroid~\cite{liu2024make} is designed to generate Android test sequences. In contrast to these works, \Name{} addresses a critical bottleneck and devises a solution that transforms the challenge of general NCF bug detection from nearly impossible to feasible.

\section{Conclusion}

Our study provides empirical evidence on the effectiveness of using Large Language Models (LLMs) as test oracles for detecting non-crash functional (NCF) bugs in Android applications. We propose \Name{}, which employs an LLM to determine whether a given test sequence reveals an NCF bug, using two tailored prompts. Our results indicate that \Name{} outperforms state-of-the-art tools in detecting Android NCF bugs. Additionally, we found that the two-prompt strategy, in-context learning, and prompt rules significantly enhance \Name{}'s performance. Among the five different LLM models tested, GPT-4o demonstrated outstanding performance compared with others. Furthermore, we show that \Name{} can uncover previously unknown functionalities in real-world applications. By highlighting some limitations of \Name{}, our study provides a foundation for future research and development, aiming to refine and enhance the effectiveness of LLM-based test oracles in Android apps.

\bibliographystyle{IEEEtran}
\bibliography{sample-base}

\end{document}